\documentclass[12pt]{article}
\usepackage{amsfonts}

\def\hybrid{\topmargin 0pt      \oddsidemargin 0pt
        \headheight 0pt \headsep 0pt
        \voffset=-0.5cm
        \textwidth 6.5in        
        \textheight 9in         
        \marginparwidth 0.0in
        \parskip 4pt plus 1pt   \jot = 1.5ex}
\catcode`\@=11
\def\marginnote#1{}

\newcount\hour
\newcount\minute
\newtoks\amorpm
\hour=\time\divide\hour by60
\minute=\time{\multiply\hour by60 \global\advance\minute by-\hour}
\edef\standardtime{{\ifnum\hour<12 \global\amorpm={am}%
        \else\global\amorpm={pm}\advance\hour by-12 \fi
        \ifnum\hour=0 \hour=12 \fi
        \number\hour:\ifnum\minute<10 0\fi\number\minute\the\amorpm}}
\edef\militarytime{\number\hour:\ifnum\minute<10 0\fi\number\minute}

\def\draftlabel#1{{\@bsphack\if@filesw {\let\thepage\relax
   \xdef\@gtempa{\write\@auxout{\string
      \newlabel{#1}{{\@currentlabel}{\thepage}}}}}\@gtempa
   \if@nobreak \ifvmode\nobreak\fi\fi\fi\@esphack}
        \gdef\@eqnlabel{#1}}
\def\@eqnlabel{}
\def\@vacuum{}
\def\draftmarginnote#1{\marginpar{\raggedright\scriptsize\tt#1}}
\def\draftlabel#1{{\@bsphack\if@filesw {\let\thepage\relax
   \xdef\@gtempa{\write\@auxout{\string
      \newlabel{#1}{{\@currentlabel}{\thepage}}}}}\@gtempa
   \if@nobreak \ifvmode\nobreak\fi\fi\fi\@esphack}
        \gdef\@eqnlabel{#1}}
\def\@eqnlabel{}
\def\@vacuum{}
\def\draftmarginnote#1{\marginpar{\raggedright\scriptsize\tt#1}}

\def\draft{\oddsidemargin -.5truein
        \def\@oddfoot{\sl preliminary draft {\tt(\jobname)}\hfil
        \rm\thepage\hfil\sl\today\quad\militarytime}
        \let\@evenfoot\@oddfoot \overfullrule 3pt
        \let\label=\draftlabel
        \let\marginnote=\draftmarginnote
   \def\@eqnnum{(\theequation)\rlap{\kern\marginparsep\tt\@eqnlabel}%
\global\let\@eqnlabel\@vacuum}  }

\def\numberbysection{\@addtoreset{equation}{section}
        \def\theequation{\thesection.\arabic{equation}}}

\def\titlepage{\@restonecolfalse\if@twocolumn\@restonecoltrue\onecolumn
     \else \newpage \fi \thispagestyle{empty}\c@page\z@
        \def\thefootnote{\fnsymbol{footnote}} }

\def\endtitlepage{\if@restonecol\twocolumn \else  \fi
        \def\thefootnote{\arabic{footnote}}
        \setcounter{footnote}{0}}  
\catcode`@=12
\relax

\numberbysection
\hybrid

\def\beq{\begin{equation}}
\def\eeq{\end{equation}}
\def\bea{\begin{eqnarray}}	
\def\eea{\end{eqnarray}}	
\def\p{\partial}
\def\G{\Gamma}
\def\g{\gamma}
\def\s{\sigma}
\def\z{\zeta}

\def\a{\alpha}
\def\b{\beta}
\def\e{\varepsilon}
\def\l{\lambda}

\def\B{{\cal B}}

\def\V{{\cal V}}

\def\F{{\cal F}}

\def\L{{\cal L}}

\def\O{{\cal O}}

\def\wh{\widehat}

\newtheorem{theo}{Theorem}[section]

\newtheorem{lem}{Lemma}[section]
\newtheorem{rem}{Remark}[section] 

\def\bC{\mathbb C}
\def\bZ{\mathbb Z}
\def\bN{\mathbb Z_{>0}}
\def\bP{\mathbb P}
\def\bt{\mathbf t}
\def\fm{\mathfrak m}
\def\Nm{\mathop{\rm Nm}\nolimits}
\def\Pic{\mathop{\rm Pic}\nolimits}
\def\Prym{\mathop{\rm Prym}\nolimits}
\def\res{\mathop{\rm res}\nolimits}
\def\bea{\begingroup\arraycolsep=1.5pt\begin{eqnarray}}
\def\eea{\end{eqnarray}\endgroup}
\def\blist{\begin{list}{}{\topsep2pt plus1pt minus1pt\itemsep1pt
plus.5pt minus.5pt\labelwidth\leftmargin\advance\labelwidth by-\labelsep}}
\def\elist{\end{list}}
\begin{document}
 \title{Abelian solutions of the KP equation}
\author{I. Krichever%
\thanks{Columbia University, New York, USA and
Landau Institute for Theoretical Physics, Moscow, Russia; e-mail:
krichev@math.columbia.edu. Research is supported in part by National Science
Foundation under the grant DMS-04-05519.}
\and T. Shiota%
\thanks{Kyoto University, Kyoto, Japan; e-mail: shiota@math.kyoto-u.ac.jp.
Research is supported in part by Japanese Ministry of Education, Culture, Sports,
Science and Technology under the Grant-in-Aid for Scientific Research (S)
18104001.}}

\date{January 3, 2008}

\maketitle

 \begin{abstract}We introduce the notion of abelian solutions of KP
 equations and show that all of them are algebro-geometric.
 \end{abstract}


\section{Introduction}
The Kadomtsev-Petviashvili equation (KP)
\begin{equation}\label{kp}
\frac{3}{4} u_{yy}=\frac{\p}{\p x} \biggl( u_t-\frac{1}{4}u_{xxx}-
\frac{3}{2} u u_x \biggr)
\end{equation}
is one of the most fundamental integrable equation of the soliton theory.
Various classes of its exact solutions have been constructed and studied over
the years. The purpose of this paper is to introduce and characterize a new
class of solutions of the KP equation. We call a solution $u(x,y,t)$
of the KP equation \emph{abelian} if it is of the form
\beq\label{u}
u=-2\p_x^2\ln \tau(Ux+z,y,t)\,,
\eeq
where $x$, $y$, $t\in\bC$ and $z\in \bC^n$ are independent variables,
$0\ne U\in\bC^n$, and for all $y$, $t$ the function
$\tau(\cdot,y,t)$ is a holomorphic section of a line bundle $\L=\L(y,t)$ on an
abelian variety $X=\bC^n/\Lambda$, i.e., for all $\l\in\Lambda$
it satisfies the monodromy relations
\begin{equation}\label{mon}
\tau(z+\l,y,t)=e^{a_\l\cdot z+b_\l}\tau(z,y,t),\quad
\hbox{for some $a_\l\in\bC^n$, $b_\l=b_\l(y,t)\in\bC$}\,.
\end{equation}

There are two particular cases in which a complete characterization of the abelian
solutions has been known for years. The first one is the case $n=1$ of
elliptic solutions of the KP equations. Theory of elliptic solutions of
the KP equation goes back to the work \cite{amkm}, where it was found that
the dynamics of poles of the elliptic (resp.\ rational or trigonometric)
solutions of
the Korteweg-de~Vries equation can be described in terms
of the elliptic (resp.\ rational or trigonometric)
Calogero-Moser (CM) system with certain constraints.
In \cite{krelkp} it was shown that when the constraints are removed
this correspondence becomes an isomorphism between the
solutions of the elliptic (resp.\ rational etc.)\ CM system and the
elliptic (resp.\ rational etc.) solutions of the KP equation.
Recall that the elliptic~CM system is a completely integrable system with
Lax representation $\dot L=[L,M]$, where $L=L(z)$ and $M=M(z)$
are $(N\times N)$ matrices depending on a spectral parameter $z\in\bC$.
The involutive integrals~$H_n$ are defined as $H_n=n^{-1}\mathop{\rm Tr}L^n$.
A function $u(x,y,t)$ which is an elliptic function
in $x$ satisfies the KP equation if and only if it has the form
\begin{equation}\label{pot}
u(x,y,t)=2\sum_{i=1}^N\wp(x-q_i(y,t))+c\,,
\end{equation}
where $\wp(q)$ is the Weierstrass $\wp$-function (\cite{c}),
and its poles $q_i$ as functions of $y$ (resp.\ $t$) satisfy the equations of
motion of the elliptic CM system, corresponding to the second Hamiltonian
$$
H_2=\frac{1}{2}\sum_{i=1}^N p_i^2 - 2 \sum_{i\ne j}\wp(q_i-q_j)
$$
(resp.\ the third Hamiltonian $H_3$).

An explicit theta-functional formula for algebro-geometric solutions
of the KP equation provides an \emph{algebraic} solution
of the Cauchy problem for the elliptic CM system \cite{krelkp}.
Namely, for generic initial data the positions $q=q_i(y,t)$ of the particles
at any time~$y$, $t$ are roots of the equation
$$
\theta \bigl(U q+V y+Wt+Z)=0\,,
$$
where $\theta(Z)$ is the Riemann theta-function of the Jacobian of
\emph{time-independent} spectral curve~$\G$,
given by $R(k,z)=\det (kI-L(z))=0$ (hence $\G$ as well as
the vectors $U$, $V$, $W$ and $Z$ depend on the initial data).

The correspondence between finite-dimensional integrable systems and
pole systems of various soliton equations has been extensively studied
in \cite{krbab,krnest,kreltoda,krwz,krzab}. A general scheme of
constructing such systems using a specific inverse problem for
linear equations with elliptic coefficients is presented in \cite{krnest}.
In \cite{akv} it was generalized for the case of field analog of the CM system
(see also \cite{loz}).

The second case in which a complete characterization of abelian solutions
is known is the case of
indecomposable principally polarized abelian variety (ppav).
The corresponding $\theta$-function is unique up to normalization, so that
Ansatz (\ref{u}) takes the form $u=-2\p_x^2\ln\theta(Ux+Z(y,t)+z)$.
Since the flows commute, $Z(y,t)$ must be linear in $y$ and $t$:
\beq\label{u1}
u=-2\p_x^2 \ln\theta(Ux+Vy+Wt+z)\,.
\eeq
Novikov conjectured that an indecomposable  ppav $(X,\theta)$ is the Jacobian of
a smooth genus $g$ algebraic curve if and only if there exist vectors
$U\mathrel(\neq 0)$, $V$ and $W$ such
that $u$ given by (\ref{u1}) satisfies the KP equation. Novikov's
conjecture was proved in \cite{shiota}, and until recently has remained the
most effective solution of the Riemann-Schottky problem.

Besides these two cases of abelian solutions with known characterization,
another may be worth mentioning.
Let $\G$ be a curve, $P\in\G$ a smooth point, and
$\pi\colon\G\to\G_0$ a ramified covering map such that the curve $\G_0$
has arithmetic genus $g_0 > 0$ and $P$ is a branch point of the covering.
Let $J(\G)=\Pic^0(\G)$ be the (generalized) Jacobian of $\G$, let
$\Nm\colon J(\G)\to J(\G_0)$ be the reduced norm map as in
\cite{mumford_prym}, and let
$$
X=\ker(\Nm)^0\subset J(\G)
$$
be the identity component of the kernel of Nm. Suppose $X$ is compact.
By assumption we have
$$
	\dim J(\G) - \dim X = \dim J(\G_0) = g_0 > 0,
$$
so that $X$ is a proper subvariety of $J(\G)$, and the polarization on $X$
induced by that on $J(\G)$ is \emph{not} principal.
Now take a local coordinate $\z\in\fm_P\setminus\fm_P^2$ at $P$,
and define the KP flows on $\overline{\Pic^{g-1}}(\G)$ using
the data $(\G,P,\z)$.

Suppose first that $\pi$ is given by $\z\mapsto\z^m$ near $P$, i.e.,
$\z^m\in\pi^*(\fm_{\pi(P)}\setminus\fm_{\pi(P)}^2)$.
Then for any $r\in\bN$ \emph{not} divisible by $m$
we have $\prod_{j=0}^{m-1}e^{t_r/(\e^j\z)^r } = 1$, where
$\e=e^{2\pi i/m}$, so that, as seen from the definition of the map
$e$ in (\ref{expmap}) below, we have
$e(0,\dots,0,t_r,0,\dots)\in X$, so the $r$-th KP orbit of
$\F\in\overline{\Pic^{g-1}}(\G)$ is contained in
$\F\otimes X:=\{\F\otimes\L\mid\L\in X\}\subset\overline{\Pic^{g-1}}(\G)$.

In general, since for any $r_0\in\bN$ the space
$\sum_{r\le r_0}\bC\p/\p t_r$ is independent of the choice of $\z$,
for any $\z\in\fm_P\setminus\fm_P^2$ and $0<r<m$
(so in particular for $r=1$), the $r$-th KP orbit of $\F$
is contained in $\F\otimes X$, and so it gives an abelian solution.
Let us call this the \emph{Prym-like} case.

An important subcase of it is the quasiperiodic solutions of
Novikov-Veselov (NV) or BKP hierarchies.
In the Prym-like case, just as in the NV/BKP case we can put
singularities to $\G$ and $\G_0$ in such a way that $X$ remains
compact, so it is more general than the KP quasiperiodic solutions,
for which $J(\G)$ itself is compact.
Recall that NV or BKP quasiperiodic solutions can be obtained from
Prym varieties $\Prym(\G,\iota)$ of curves $\G$ with involution $\iota$
having two fixed points.
The Riemann theta function of $J(\G)$ restricted to a suitable translate
of $\Prym(\G,\iota)$
becomes the square of another holomorphic function, which defines the
principal polarization on $\Prym(\G,\iota)$.
The Prym theta function becomes NV or BKP tau function, whose square
is a special KP tau function with all \emph{even} times set to zero,
so any KP~time-translate of it
\begin{itemize}\itemsep0pt
\item gives an abelian solution of the KP hierarchy
with $n=\dim X$ being one-half the genus $g(\G)$ of $\G$, and
\item defines twice the principal polarization on $X$.
\end{itemize}
A natural question may be whether these conditions characterize
the (time-translates of) NV or BKP quasiperiodic solutions.

Hurwitz' formula tells us that in the Prym-like case
$n=\dim(X)\ge g(\G)/2$, where the equality holds only in
the NV/BKP case.  At the moment we have no examples of
abelian solutions with $1<n<g(\G)/2$.

\paragraph{Integrable linear equations}
The KP equation can be seen as the compatibility condition for a
system of linear equations.  In \cite{kr-schot}, it is shown that
only one of the \emph{auxiliary} linear equations
\beq\label{lax1}
\left(\p_y-\p_x^2+u\right)\psi=0
\eeq
suffices to characterize the Jacobian locus. We shall call this an
\emph{intebrable linear equation} although here both $u$ and $\psi$
are unknown, and the equation should be regarded nonlinear.

This result is stronger than the one given in terms of the KP equation
(see details in \cite{flex}).
In terms of the Kummer map it is equivalent to the characterization of
the Jacobians via flexes of the Kummer varieties, which is one out of
the three particular cases of the trisecant conjecture, first formulated
in \cite{wel1}.
Two remaining cases of the trisecant conjecture were proved in \cite{kr-tri}.
The characterization problem of the Prym varieties among indecomposable ppav
was solved in \cite{kr-quad, kr-prym}.

The notion of abelian solutions can be extended to equation (\ref{lax1}).
A solution $(u,\psi)$ of equation (\ref{lax1}) is \emph{abelian} if
\beq\label{u2}
u=-2\p_x^2 \ln \tau (Ux+z,y) \quad{\rm and}\quad
\psi={\tau_A(Ux+z,y)\over\tau(Ux+z,y)}\, e^{p\,x+E\,y}
\eeq
for some $p$, $E\in\bC$ and $0\ne U\in\bC^n$, such that
$\tau_A(z,y)$ and $\tau(z,y)$ are holomorphic functions of
$(z,y)\in\bC^n\times D$, where $D$ is a neighborhood of 0 in $\bC$,
satisfying the monodromy properties
\beq\label{monod}
\tau(z+\l,y)=e^{a_\l\cdot z+b_\l(y)}\tau(z,y)\,,\quad
\tau_A(z+\l,y)=e^{a_\l\cdot z+c_\l(y)}\tau_A(z,y)
\eeq
for all $\l$ in the period lattice $\Lambda$ of
an abelian variety $X=\bC^n/\Lambda$.

\paragraph{Main result}
Our examples of abelian solutions of KP equation (\ref{kp}) or the
integrable linear equation (\ref{lax1}) can be extended to
rank one algebro-geometric solutions of the KP hierarchy, for which
$X\subset J(\G)$, with $\G$ being the spectral curve.
In this paper we follow the lines of \cite{kr-schot} to
observe that abelian solutions of (\ref{kp}) or (\ref{lax1}) are
rank one algebro-geometric, and $X\subset J(\G)$ holds if
the group $\bC U=\{Ux\in X\mid x\in\bC\}$ is Zariski dense in $X$.

Without loss of generality it will be assumed throughout the paper that
\blist
\item[$(*)$]$\Lambda$ is a maximal lattice satisfying the respective
monodromy property, i.e., any $\l\in\bC^n$ which satisfies
condition (\ref{mon}) in the case of KP equation (\ref{kp}), or
condition (\ref{monod}) in the case of equation (\ref{lax1}),
must belong to $\Lambda$.
\elist
\begin{theo}\label{main1} Suppose that one of the following two conditions
{\rm (A)}, {\rm (B)} holds:
\blist
\item[\rm(A)]for any
$z\in\bC^n$, and $y$, $t$ in a neighborhood of the origin in $\bC^2$,
the function $u$ given by (\ref{u}), with $\tau$ satisfying the
monodromy condition (\ref{mon}), is an abelian solution of
the KP equation (\ref{kp});
\item[\rm(B)]for any
$z\in\bC^n$, and $y$ in a neighborhood of the origin in $\bC$,
the pair $(u,\psi)$ given by (\ref{u2}), with $\tau$ and $\tau_A$
satisfying the monodromy condition (\ref{monod}), is an abelian solution of
equation (\ref{lax1}), such that the following condition holds:
\blist
\item[$(\dag)$]
\label{condition-dag}the divisors
$\Theta:=\{(z,y)\in X\times D\mid\tau(z,y)=0\}$ and
$\Theta_A:=\{(z,y)\in X\times D\mid\tau_A(z,y)=0\}$ have
no common component.
\elist
\elist
Suppose, moreover, that condition $(*)$ holds. Then there exist
a unique irreducible algebraic curve $\G$,
a smooth point $P\in\G$,
a subabelian variety $Y$ of $X$ containing $\bC U$,
where $u$ is as in (\ref{u}) or (\ref{u2}),
an injective homomorphism $i\colon Y\hookrightarrow J(\G)$,
a $Y$-invariant blow-up $\pi\colon\tilde X\to X$ (i.e., the $Y$-action
on $X$ lifts to $\tilde X$) with the center contained in
\beq\label{Sig}
\Sigma:=\bigcap_{x\in\bC}(\Theta+Ux)\,,
\eeq
and
a holomorphic map $j$ of $\tilde X$ to the space $\overline{\Pic^{g-1}}(\G)$
of torsion-free rank 1 sheaves on $\G$ of degree $g-1$,
where $g=g(\G)$ is the arithmetic genus of $\G$,
such that for any given $\tilde z\in \tilde X$ the diagram
\beq\label{is}
\begin{array}{ccc@{}lcc@{}c@{}c}
\bC&\stackrel{e_1}{\longrightarrow}&
Y&{}\ni z'&\longmapsto&z'+\tilde z&{}\in{}&\tilde X
\\
&&\Big\downarrow\vcenter{\rlap{$\scriptstyle i$}}
&&&&&\Big\downarrow\vcenter{\rlap{$\scriptstyle j$}}
\\[4pt]
&&J(\G)&{}\ni\L&\longmapsto&\L\otimes j(\tilde z)&{}\in{}&
\overline{\Pic^{g-1}}\rlap{$(\G)$}
\end{array}
\eeq
commutes,
where $e_1(x)=Ux\in Y$, and such that, locally in $\tilde z\in\tilde X$,
\beq\label{is1}
\tau(Ux+z,y,t)=\rho(\tilde z,y,t)\,\wh\tau(x,y,t,0,\ldots\mid\G,P,j(\tilde z))
\eeq
(here the $t$-variable is absent in case $(B)$), where $z=\pi(\tilde z)$,
$\wh\tau(t_1,t_2,t_3,\ldots \mid \G,P,\F)$ is the KP tau-function
defined by the data $(\G,P,\F)$, and $\rho(\tilde z,y,t)\not\equiv0$
is a function of $(\tilde z,y,t)$ which satisfies $\p_U\rho=0$.
\end{theo}
Here are some remarks:
\blist
\item[$\bullet$]
the locus $\Sigma$, defined in (\ref{Sig}), is a unique maximal
$\p_U$-invariant subset of $\Theta$, and it will be called the
\emph{singular locus},
\item[$\bullet$]
the main assumptions in either case (A) or (B), i.e., (\ref{kp}) or
(\ref{lax1}), contain excessive information.
All what is used for their proof is a certain equation valid on the
$\tau$-divisor derived in Lemmas~3.1 and 3.2 below.
\item[$\bullet$]
time evolutions of the KP hierarchy can be described by
extending the map $e_1$ in (\ref{is}):
$$
\begin{array}{ccccc}
\bC&\stackrel{e_1}{\longrightarrow}&
Y&\stackrel{\cdot\,+\,\tilde z}{\lhook\joinrel\longrightarrow}&\tilde X
\\
\Big\downarrow\vcenter{\rlap{$\scriptstyle j_1$}}&&
\Big\downarrow\vcenter{\rlap{$\scriptstyle i$}}&&
\Big\downarrow\vcenter{\rlap{$\scriptstyle j$}}
\\
\bC^\infty&\stackrel{e}{\longrightarrow}&
J(\G)&\stackrel{\cdot\,\otimes\,j(\tilde z)}{\longrightarrow}&
\overline{\Pic^{g-1}}(\G)\rlap{\,,}
\end{array}
\eqno(\ref{is}')
$$
where $j_1\colon x\mapsto(x,0,0,\dots)$, and by taking a local coordinate
$\z\in\fm_P\setminus\fm_P^2$ at $P$,
the homomorphism $e$ is defined by
\beq\label{expmap}
e(t_1,t_2,\dots)=\left\{\begin{tabular}{@{\,}l}
$\O$ near $P$ and on $\G\setminus\{P\}$,\\
glued to itself around $P$ by $e^{\sum t_i/\z^i}$\,.
\end{tabular}\right.
\eeq
\item[$\bullet$]
the factor $\rho$ in (\ref{is1}) is needed
since multiplying $\tau$ and $\tau_A$ by
a quantity independent of $x$ has no effect on (\ref{u}) or (\ref{u2}).
\elist
Since $i$ and $e$ are homomorphisms, they are lifted to linear maps on the
universal coverings, as readily seen for the latter in the formula for tau
when $\G$ is smooth:
\beq\label{is2}
\wh\tau(x,t_2,t_3,\dots\mid\G,P,j(\tilde z))=
\theta\Bigl(Ux+\sum V_it_i+j(\tilde z)\Bigm|B(\G)\Bigr)\,
e^{Q(x,t_2,t_3,\ldots)}\,,
\eeq
where $V_i\in\bC^n$, $Q$ is a quadratic form, $B(\G)$ is the matrix of
$B$-periods of $\G$, and $\theta$ is the Riemann theta function.
This linearization of nonlinear $t_i$-dynamics provides some
evidence that there might be underlying integrable systems on the spaces of
higher level theta-functions on an abelian variety.
The CM system is an example of such a system for $n=1$.

\paragraph{Blow-up and $\bP^1$-family of solutions}
The space of tau functions is the total space, say $\B$, of a
$\bC^*$-bundle over $\overline{\Pic^{g-1}}(\G)$.  However, given
$z\in X$ our $\tau$ as a function of $x$, $y$, $t$ (or $\wh\tau$
as a function of $t_1$, $t_2$, \dots) might be identically zero.
So this maps $X$ to $\B\cup\{0\}$, a space to which the projection from
$\B$ to $\overline{\Pic^{g-1}}(\G)$ cannot be continuously extended.
Thus we often do need to blow up $X$ to define $j$ in (\ref{is}).
However, we observe
\begin{rem}
No blow-up is needed if $\G$ is smooth.
\end{rem}
\emph{Proof.} After dividing $\tau$ by the trivial factors (see
Section~\ref{sec:zeroloc}), we assume
the locus $\Sigma$ is of codimension${}\ge2$ in $X$.

Suppose $\G$ is smooth.  Then $\overline{{\rm Pic}^{g-1}}(\G)=J(\G)$,
and it is an abelian variety.  Then any holomorphic map from $\bP^1$
to it must be constant. Indeed, since $\bP^1$ is simply connected, any
such map can be lifted to a map from $\bP^1$ to the universal covering
of $J(\G)$, i.e., an affine space.  Hence it must be constant.

Assuming $\Sigma\ne\emptyset$, take any point $p_0\in\Sigma$, and
take a 2-dimensional plane $\Pi\subset X$ such that locally near $p_0$,
the loci $\Sigma$ and $\Pi$ meet only at $p_0$.  Take a coordinate
system $(a,b)$ on $\Pi$ such that $p_0$ is the origin $a=b=0$,
and restrict the range of $z$ to $\Pi$ to obtain a family of $\tau$
parametrized by $(a,b)$.  Taylor expanding $\tau$ in $a$, $b$:
$$
\tau(x,y;a,b)=\sum_{m,n\ge0}\tau_{m,n}(x,y)a^m b^n ,
$$
where we omit the $t$-variable in case (A) (or the sequence
$t_3$, $t_4$, \dots if $\tau$ is a KP $\tau$-function),
let $N$ be the set of indices $(m,n)$ for which $\tau_{m,n}\not\equiv0$.
Since $\tau(x,y,0,0)\equiv 0$ and
$\tau$ is not divisible by $a$ or $b$, we have $(0,0)\notin N$, and
$(m,0)\in N$, $(0,n)\in N$ for some $m$, $n>0$.
Hence there exist positive integers $p$, $q$ and $C$ such that
$N\subset\{(m,n) \mid pm + qn \ge C\}$, and such that $N$ meets the
line $pm + qn = C$ at least at two points.  Then as the ``lowest
order'' part of $\tau$,
$$
\tilde\tau(x,y;a,b):=\sum_{pm+qn=C}\tau_{m,n}(x,y)a^m b^n
$$
is a family of solutions, and it is a weighted homogeneous polynomial
of $a$ and $b$.  Then $\tilde\tau(x,y;a^p,b^q)$ is
an (unweighted) homogeneous polynomial, giving a $\bP^1$-family of
$\tau\bmod\bC^\times$.  Then by the fact noted above, this must be a constant
family, so all the $\tau_{m,n}$ on the line $pm+qn=C$ must be a
constant multiple of the same $\tau$.  Observing this on every edge
of the polygon $N$, we see that as $(a,b)\to(0,0)$ we have a
well-defined limit of the corresponding sheaf $\cal F$, which means
no blow-up is necessary around $p_0$.  Since the point $p_0\in\Sigma$ and
the plane $\Pi\ni p_0$ are arbitrary, no blow-up is needed at all, so
the remark follows.
\medskip


Having this in mind, let us start with a curve with a node, and constract
a nontrivial family of $\tau$-functions of the form
\beq\label{P1fam}
\tau(\bt,z';a,b) = a \tau_0(\bt,z') + b \tau_1(\bt,z')\,,\quad
(a,b)\in\bC^2\setminus\{(0,0)\}\,,
\eeq
where $\tau_0$ and $\tau_1$ (and hence the entire family) depend on
the same parameters $z'\in\bC^d$ in such a way that
$\tau_i(\bt,z')=\tau_i(z'+Ut_1,t_2,\dots)$, and satisfy the same
monodromy conditions for a lattice $\Lambda'\subset\bC^d$.

Such a family of quasiperiodic $\tau$-functions should yield an example
of abelian solutions for which blow-up is really needed:
take an abelian variety $Z=\bC^n/\Lambda$, and two $\bC$-linearly
independent functions $\theta_0$ and $\theta_1$ on $\bC^n$
which satisfy the same monodromy conditions with respect to $\Lambda$
(so the ratio $\theta_0/\theta_1$ is a meromorphic function on $Z$),
let $Y=\bC^d/\Lambda'$, $X=Y\times Z$, denote $X\ni z=(z',z'')$,
with $z'\in Y$ and $z''\in Z$, and define $\tau(\bt,z)$ by replacing
$a$ and $b$ in (\ref{P1fam}) by $\theta_0(z'')$ and $\theta_1(z'')$,
respectively, i.e.,
$$
\tau(\bt,z',z'')=\theta_0(z'')\tau_0(\bt,z')+
\theta_1(z'')\tau_1(\bt,z')\,.
$$
We need to blow up $X$ along the intersection of zero loci of
$\theta_0$ and $\theta_1$ to define a map to
$\overline{{\rm Pic}^{g-1}}(\G)$, where $g$ is the
arithmetic genus of $\G$.  Note also that the KP hierarchy has
no control over the 2nd factor $Z$.

Construction of family (\ref{P1fam}) goes as follows.  First, consider
a simple Backlund transform applied to any quasiperiodic $\tau$-function
$\tau_0(\bt)$, $\bt=(t_1,t_2,\dots)$.  This yields a family of
$\tau$-functions of the form (\ref{P1fam}), where
\beq\label{DJKMvertex}
\tau_1:=X(p,q)\tau_0:=\exp\Bigl(\sum t_i(p^i-q^i)\Bigr)
\exp\Bigl(\sum\frac{q^{-i}-p^{-i}}{i}\frac{\p}{\p t_i}\Bigr)
\tau_0
\eeq
using Date et al.'s notation for vertex operator \cite{djkm}.
It is more common to take $a=1$, but formula (\ref{P1fam}) gives a
tau function as long as $(a,b)\ne(0,0)$.
Let us try to make $\tau_0$ and $\tau_1$ satisfy the same monodromy
conditions.  If $\tau_0$ is a quasiperiodic solution
associated with a smooth curve $\tilde\G$ and a point
$P\in\tilde\G$, the effect of $a + bX(p,q)$ on $\tau_0$ is
to identify the points $p$ and $q$ on $\tilde\G$ to make a curve $\G$
with node, and the fibres of line bundle on
$\tilde\G$ at $p$ and $q$ to obtain a line bundle on $\G$ if the
ratio $b/a$ is not 0 or $\infty$, or a torsion-free rank 1 sheaf
on $\G$ in general.

As we saw in the paragraph on Prym-like solutions, suitably chosen
$\tilde\G$, $P$, $p$ and $q$ make entire family (\ref{P1fam}) of
$\tau$-functions quasiperiodic in $t_1$: suppose there exists a
ramified covering map $\tilde\pi$ of $\tilde\G$ to another smooth curve
$\tilde\G_0$ of genus $g_0\ge0$, such that $P$, $p$ and $q$ are branch
points, and such that $\tilde\pi^{-1}(\tilde\pi(p))=\{p\}$ and
$\tilde\pi^{-1}(\tilde\pi(q))=\{q\}$ hold. 
Identify $p$ and $q$ on $\tilde\G$, and
$\tilde\pi(p)$ and $\tilde\pi(q)$ on $\tilde\G_0$
to obtain curves with nodes $\G$ and $\G_0$, respectively, with a
covering map $\pi\colon\G\to\G_0$.  Note that $p_a(\G_0)=g_0+1\ge1$.
Since $J(\G)$ (resp.\ $J(\G_0)$) is a $\bC^\times$-extension of
$J(\tilde\G)$ (resp.\ $J(\tilde\G_0)$), and since the restriction of
$\Nm\colon J(\G)\to J(\G_0)$ to the $\bC^\times$ does not vanish,
the identity component $Y$ of $\ker(\Nm)$ is
an abelian variety isogenous to that of
$\ker\bigl(\widetilde{\Nm}\colon J(\tilde\G)\to J(\tilde\G_0)\bigr)$.
Hence, as seen in our construction of ``Prym-like'' solutions,
the $t_1$-evolution associated to $(\G,P)$ is contained in $Y$
and hence quasiperiodic, and for any $(a,b)\ne(0,0)$
the solution $\tau$ in (\ref{P1fam}) is quasiperiodic in $t_1$.

Next, let us construct a more explicit example.
Starting with an elliptic curve $\tilde\G$, we can easily adjust
the monodromy conditions of $\tau_0$ and $\tau_1$ so that, after
doubling one of the fundamental periods (or replacing
$J(\tilde\G)\simeq\tilde\G$ by a double cover, $Y$, of it),
$\tau_0$ and $\tau_1$ satisfy the same monodromy conditions.
In this example we take $\tilde\G_0=\bP^1$, so that
$\ker(\widetilde{\Nm})=J(\tilde\G)=\tilde\G$.
That $Y$ is a double cover of it fits the general picture above.

For brevity we restrict ourselves to the first three time variables
$(t_1,t_2,t_3)=(x,y,t)$, and consider the first KP equation (\ref{kp})
only. It is a simple exercise on elliptic functions to work out the
formulae for the whole KP hierarchy.

For $2\omega_1$, $2\omega_3\in\bC^\times$ with $\Im(\omega_3/\omega_1)>0$,
let $\Lambda_0:=2\bZ\omega_1+2\bZ\omega_3$ and $\tilde\G:=\bC/\Lambda_0$.
Denote by $\sum'$ (resp.\ $\prod'$) the sum (resp.\ product)
over all $\omega\in\Lambda_0\setminus\{0\}$.  Defining
Weierstrass' $\s$-function by
$$
\s(z):=z\prod\nolimits'(1-z/\omega)\exp(z/\omega+(z/\omega)^2/2)
$$
and using a well-known differential equation for $\wp(z)=-(\ln\s(z))''$,
we have
$$
\frac{D_z^4\s\cdot\s}{\s^2}\equiv
2\p_z^4\ln\s+12(\p_z^2\ln\s)^2=
g_2:=60\sum\nolimits'\omega^{-4}\,.
$$
Hence $\tau_0(x,y,t):=\tau_0(x,y,t,z):=e^{\a xt+\b y^2}\s(z+x)$
is a $z$-dependent solution to the KP equation iff
\beq\label{KPsigma}
g_2+12\b-8\a=0.
\eeq
Moreover, setting $\zeta(z)=(\ln\s(z))'$,
$\omega_2:=-\omega_1-\omega_3$ and $\eta_\nu=\zeta(\omega_\nu)$,
where $\nu=1$,~2,~3, we have
$$ 
\s(z+2\omega_\nu)=-\s(z)\exp(2\eta_\nu(z+\omega_\nu))
$$ 
and
$$
\left|\begin{array}{@{}c@{\ \ }c@{}}\eta_1&\eta_3\\\omega_1&\omega_3
\end{array}\right|=
\left|\begin{array}{@{}c@{\ \ }c@{}}\eta_3&\eta_2\\\omega_3&\omega_2
\end{array}\right|=
\left|\begin{array}{@{}c@{\ \ }c@{}}\eta_2&\eta_1\\\omega_2&\omega_1
\end{array}\right|=\frac{\pi i}2,
$$
so that Weierstrass' co-sigma functions
$\s_\mu(z):=\exp(-\eta_\mu z)\s(z+\omega_\mu)/\s(\omega_\mu)$, $\mu=1$,~2, 3,
satisfy
$$
\s_\mu(z+2\omega_\nu)=(-1)^{\delta_{\mu,\nu}}
\exp(2\eta_\nu(z+\omega_\nu))\s_\mu(z),\quad\nu=1,2,3,
$$
i.e., $\sigma_\mu$ satisfies the same monodromy conditions as $\sigma$
for the periods $2\omega_\mu$ and $4\omega_\nu$ ($\nu\ne\mu$).
On the other hand, (\ref{DJKMvertex}) implies
\beq\label{tauratio}
\frac{\tau_1}{\tau_0}=C_{y,t}\exp(Ax)
\frac{\sigma(x-1/p+1/q)}{\sigma(x)},
\eeq
where $A:=p-q+\a(-1/p^3+1/q^3)/3$, and $C_{y,t}\ne0$ is independent of $x$.
Therefore, if we choose $p$, $q$ and $\a$ so that
\beq\label{KPvm}
-\frac1p+\frac1q=\omega_\mu\quad\hbox{and}\quad
p-q+\frac\a3\biggl(-\frac1{p^3}+\frac1{q^3}\biggr)=2\eta_\mu
\eeq
hold for some $\mu\in\{1,2,3\}$, then the right-hand side of
(\ref{tauratio}) becomes $\s_\mu/\s$ up to a factor independent of $x$,
so that $\tau_0$ and $\tau_1$ satisfy the same monodromy conditions 
for the periods $2\omega_\mu$ and $4\omega_\nu$ ($\nu\ne\mu$).
We thus constructed a $\bP^1$-family of solutions quasiperiodic
with respect to the lattice $2\bZ\omega_\mu+4\bZ\omega_\nu$, where
$\nu\in\{1,2,3\}\setminus\{\mu\}$ is arbitrary.
\medskip

The paper is organized as follows.  In Sect.~\ref{sec:zeroloc} we show
basic properties of the zero loci of $\tau$ and $\tau_A$, and observe
the nature of condition ($\dag$).
In Sect.~\ref{sec:constr} we construct a formal wave function,
which is used in Sect.~\ref{sec:commDO} to obtain commuting
differential operators.
Almost till the very end the proof of Theorem~1.1 goes along
the lines of \cite{shiota} (in case (A)) or \cite{kr-schot} (in case (B)).
Constructing a wave function is easier in case (A) than in
case (B), and the rest of the proof is the same for both cases, so
in what follows we mainly consider case (B).

\section{Zero loci of $\tau$ and $\tau_A$}\label{sec:zeroloc}

Before constructing the wave function, let us observe some properties
of the zero loci of $\tau$ and $\tau_A$.

For a constant coefficient polynomial $P(\xi,\eta,\dots)$,
Hirota's bilinear differential operator $P(D)=P(D_x,D_y,\dots)$ is
defined by
$$
P(D)f\cdot g:=P(\p_{x'},\p_{y'},\dots)f(x+x',y+y',\dots)g(x-x',y-y',\dots)
\Bigr|_{x'=y'=\cdots=0}\,.
$$
Putting (\ref{u2}) into the left-hand side of (\ref{lax1}) and using
$$
\p_y\frac{\tau_A}{\tau}=\frac{D_y\tau_A\cdot\tau}{\tau^2}\,,\quad
\p_x\frac{\tau_A}{\tau}=\frac{D_x\tau_A\cdot\tau}{\tau^2}
\quad\hbox{and}\quad
\p_x^2\frac{\tau_A}{\tau}=\frac{D_x^2\tau_A\cdot\tau}{\tau^2}
-2\frac{\tau_A}{\tau}\frac{D_x^2\tau\cdot\tau}{\tau^2}\,,
$$
we have
$e^{-px-Ey}\tau^2(\p_y-\p_x^2+u)\psi=((D_y+E)-(D_x+p)^2)\tau_A\cdot\tau$,
so (\ref{lax1}) is equivalent to
\beq\label{hirota}
((D_y+E)-(D_x+p)^2)\tau_A\cdot\tau=0\,.
\eeq
This readily shows the symmetry $(x,y)\leftrightarrow(-x,-y)$,
$\tau\leftrightarrow\tau_A$ of equation (\ref{lax1}), and it is also handy
to find the possible forms of common factors of $\tau$ and $\tau_A$:
\begin{lem}
If
\beq\label{common}
\tau(x,y)=(x-x(y))^b\varphi\quad\hbox{and}\quad
\tau_A(x,y)=(x-x(y))^a\varphi_A,
\eeq
where $\varphi$, $\varphi_A\ne0$ at $x=x(y)$, then
\beq\label{allowedexp}
a=\frac{\nu(\nu+1)}2\,,\qquad b=\frac{\nu(\nu-1)}2\,,
\eeq
for some $\nu\mathrel(=a-b)\in\bZ$.
Conversely, for any $\nu\in\bZ$ there is a solution of (\ref{hirota})
of the form (\ref{common}) with $a$, $b$ given by (\ref{allowedexp}).
\end{lem}
Indeed, putting (\ref{common}) into the left-hand side of (\ref{hirota})
yields
$$
-D_x^2\tau_A\cdot\tau+\cdots=
-C_{ab}(x-x(y))^{a+b-2}\varphi_A\varphi+O((x-x(y))^{a+b-1})
$$
with $C_{ab}=(a-b)^2-(a+b)$, which vanishes iff (\ref{allowedexp}) holds.
Conversely, for any holomorphic solution of the heat equation
$$
f_y=f_{xx}\,,
$$
e.g., $f\in\exp(y\p_x^2)\bC[x]$, the pair
$$
\tau_A=e^{-px-Ey}f(x,\nu y)^a\quad\hbox{and}\quad\tau=f(x,\nu y)^b
$$
give a solution of (\ref{hirota}) of the form (\ref{common}).
This proves the lemma.
\medskip

We have $|\nu|\le1$ if $\psi$ is a KP wave function evaluated at a
finite value of spectral parameter $k$, so a nonempty zero locus
with higher $|\nu|$ is an obstruction for the extension problem
to be discussed in Sect.~\ref{sec:constr}.
Rather than trying to see what the occurrence of ($|\nu|>1$)-locus means
to quasiperiodic solutions, in this paper we will simply choose to exclude
these cases by assuming condition ($\dag$) in p.~\pageref{condition-dag},
or any one of the following:
\blist
\item[($\dag'$)]\label{cdn:dagprime}
$\psi$ generically has a simple pole along $\Theta\setminus\Sigma$;
\item[($\ddag$)]
$\Theta$ and $\Theta_A$ are reduced, i.e., the zeros of $\tau$
and $\tau_A$ are generically simple;
\item[($\ddag'$)]
$\Theta$ and $\Theta_A$ are irreducible;
\item[($\ddag''$)]
$\Theta$ or $\Theta_A$ is reduced and irreducible.
\elist
Indeed, we have
\begin{lem}
For a solution of (\ref{lax1}), conditions $(\dag)$, $(\dag')$
and that $|\nu|\le1$ on all components of $\Theta\setminus\Sigma$, are all
equivalent, and if the solution is quasiperiodic, then they follow from
any one of $(\ddag)$, $(\ddag')$ and $(\ddag'')$.
\end{lem}
\emph{Proof.} Since $a$ and $b$ are positive (and one of them is
greater than 1) when $|\nu|>1$, the first assertion is almost obvious.
To be precise, we have to see that $\nu$ is constant on each component
of $\Theta$ or $\Theta_A$, or at least that the $|\nu|>1$ case cannot deform
into the $|\nu|\le1$ case, i.e., there is no parameter-dependent
solution $(\tau(x,y,\z),\tau_A(x,y,\z))$ of (\ref{hirota}) which looks
like (\ref{common}) with $a$,~$b>0$ when the parameter $\z=0$ but
not when $\z\ne0$.
Indeed, such a deformation would imply that when $\z$ is close to 0
there are $b$ simple zeros of $\tau$ (and $a$ simple zeros of $\tau_A$)
staying arbitrarily close to each other in a fixed interval of $y$.
Since $a$ or $b$ must be greater than 1, we see, even without calculations,
that this is unlikely from the usual, CM-like particle system
interpretation of the motion of zeros of $\tau$ when $|\nu|\le1$.

That $(\ddag)$ implies $|\nu|\le1$ is also equally obvious (the two
conditions are equivalent if $\Sigma$ is of codimension${}\ge2$ in $X$).

Next, if $\Theta$ and $\Theta_A$ are irreducible and $|\nu|>1$ somewhere,
then we have the same $\nu$ (and the same, unequal and positive $a$ and $b$)
all over $\Theta$ and $\Theta_A$ which have the same underlying set.
Then $\tau$ and $\tau_A$ cannot define the same polarization on $X$
which contradicts (\ref{monod}).  Hence condition ($\ddag'$) also excludes
the possibility of having $|\nu|>1$.

Criterion ($\ddag''$) may be useful since it involves only $\tau$
(or $\tau_A$).  If, e.g., $\Theta$ is reduced, then $b\le1$, and the
only case with $|\nu|>1$ we can have is $\nu=2$ ($a=3$, $b=1$).
If, moreover, $\Theta$ is irreducible, then as a divisor
$\Theta_A\ge3\Theta$, so again $\tau$ and $\tau_A$ cannot
define the same polarization on $X$.
This completes the proof of the lemma.\medskip

Thus, in what follows we assume $|\nu|\le1$, so
$\tau$ and $\tau_A$ have no common factor depending on $x$, and
$\psi$ has a simple pole along $\Theta\setminus\Sigma$.
The latter form of the condition will be used in Sect.~\ref{sec:constr}.

A pair $(\tau,\tau_A)$ can also have ``trivial'' common factors.
If $(\tau_0(x,y),\tau_{A0}(x,y))$ solves equation (\ref{hirota}),
then so does
\beq\label{snapshot}
(\rho(y)\tau_0(x,y),\rho(y)\tau_{A0}(x,y))
\eeq
for any $\rho(y)$.
Such a factor $\rho(y)$ itself is harmless, but it may not if, e.g., it
deforms into an $x$-\emph{dependent} factor in a solution with parameters.
So let us introduce a parameter $z'$, and prove that a family of solutions
of (\ref{hirota}) which is a deformation of the pair in (\ref{snapshot})
must be of the form
\beq\label{defo}
(\rho(y,z')\tau_0(x,y;z'),\rho(y,z')\tau_{A0}(x,y;z'))\,,
\eeq
where $(\tau_0(x,y;z'), \tau_{A0}(x,y;z'))$ is a family of solutions
of (\ref{hirota}) and $\rho(y,z')$ is a function of $(y,z')$
(independent of $x$) such that
$$
(\tau_0(x,y;0),\tau_{A0}(x,y;0))=(\tau_0(x,y),\tau_{A0}(x,y))
\quad\hbox{and}\quad
\rho(y,0)=\rho(y)\,.
$$
As in the KP case, such a factorization is not free, but it can be
proved using quasiperiodicity:
\begin{lem}\label{lemB}
Let $D$ and $D'$ be neighborhoods of $0$ in $\bC$, let $d\in\bN$,
let $\Lambda$ be a lattice in $\bC^d$, and let $U\in\bC^d$ be such that
$\bC U\bmod\Lambda$ is Zariski dense in $Y:=\bC^d/\Lambda$.
Let $(\tau,\tau_A)$ be a pair of functions defined on
$\bC^d\times D\times D'$, such that
\blist
\item[i)]as a pair of functions of $(x,y)\in\bC\times D$,
$(\tau(z+Ux,y,z'),\tau_A(z+Ux,y,z'))$ solves (\ref{hirota}),
\item[ii)]
$\tau$ and $\tau_A$ satisfy the same monodromy conditions in $z$:
for all $\l\in\Lambda$, there exist $a_\l\in\bC^d$, $b_\l(y,z')\in\bC$
such that
$$
\tau(z+\l,y,z')=e^{a_\l\cdot z+b_\l(y,z')}\tau(z,y,z')\,,\quad
\tau_A(z+\l,y,z')=e^{a_\l\cdot z+b_\l(y,z')}\tau_A(z,y,z')\,,
$$
\item[iii)]
$\tau(z,y,0)=y^m\tau_0(z,y)$, $\tau_A(z,y,0)=y^m\tau_{A0}(z,y)$ for some
functions $\tau_0$, $\tau_{A0}$ and $m\in\bN$.
\elist
Then the whole family $(\tau,\tau_A)$ must be factored as in (\ref{defo}),
i.e., there exist a pair of functions $(\tau_0,\tau_{A0})$ defined on
$\bC^d\times D\times D'$ and a function $\rho$ defined on $D\times D'$
such that $\rho(y,0)=y^m$,
$\tau_0(z,y,0)=\tau_0(z,y)$ and $\tau_{A0}(z,y,0)=\tau_{A0}(z,y)$, and
such that the factoring in iii) extends to $z'$ away from $0$:
$$
\tau(z,y,z')=\rho(y,z')\tau_0(z,y,z')\,,\quad
\tau_A(z,y,z')=\rho(y,z')\tau_{A0}(z,y,z')\,.
$$
\end{lem}
This is how we get the $\rho$ in (\ref{is1}).
Our choice of the factor $y^m$ in iii) is for notational simplicity only.
One can replace it by a more general $\rho_0(y)$.

Dividing $\tau$ and $\tau_A$ by the trivial factors, we may assume that
$\Sigma$ and
\beq\label{SigmaA}
\Sigma_A:=\bigcap_{x\in\bC}(\Theta_A+Ux)
\eeq
are of codimension $\ge2$ in $X\times D$.
Then we can prove that $\Sigma=\Sigma_A$, and that $\Sigma$ is
not only $\p_U$-invariant but also $\p_y$-invariant. 

Note that the division by trivial factors may change the monodromy condition
(\ref{monod}), but it will not affect our argument in the following sections
since the trivial factors $\rho$, $\l_1$ and $\l_2$ must be constant
in the directions of the Zariski closure $Y_U$ of line $\bC U$ in $X$.

\section{Construction of the wave function}\label{sec:constr}

In the core of the proof of Theorem is the construction of
quasiperiodic wave function as in (\ref{ps}) below.  Having a spectral
parameter $k$, it contains much more information than the function
$\psi$ in (\ref{u2}).  Taking (\ref{lax1}) as
a starting point, we closely follow the argument from the beginning of
Section~2 through Lemma~3.2 of \cite{kr-schot}.
The construction is presented in two steps, first locally to show that
(\ref{lax1}) guarantees the single-valuedness of the wave function
around each \emph{simple} zero in $x$ of $\tau(x,y)$,
and then globally to maintain quasiperiodicity.

\paragraph{Step 1}
Let $\tau(x,y)$ be a holomorphic function of the variable $x$ in some
domain in $\bC$, depending smoothly on a parameter $y$ and having only
simple zeros at $x=x_i(y)$:
\beq\label{xi}
\tau(x_i(y),y)=0,\quad \tau_x(x_i(y),y)\neq 0.
\eeq
Let $v_i$ and $w_i$ be the second and the third Laurent coefficients of
$u(x,y)=-2\p_x^2\ln\tau(x,y)$ at $x=x_i$, i.e.,
\beq\label{e1}
u(x,y)={2\over (x-x_i(y))^2}+v_i(y)+w_i(y)(x-x_i(y))+\cdots\,.
\eeq
\begin{lem}[\cite{flex}]
If equation (\ref{lax1}) with the potential
$u=-2\p_x^2\ln \tau(x,y)$ has a meromorphic solution $\psi_0(x,y)$, then
the equations
\beq\label{cm5}
\ddot x_i=2w_i
\eeq
hold, where the ``dots" stand for $y$-derivatives.
\end{lem}
\emph{Proof.}
Consider the Laurent expansion of $\psi_0$ at $x=x_i$:
\beq\label{psie}
\psi_0={\a_i\over x-x_i}+\b_i+\g_i(x-x_i)+\delta_i(x-x_i)^2+O((x-x_i)^3)\,.
\eeq
All coefficients in this expansion are smooth functions of the variable $y$,
and $\a_i\not\equiv0$ due to condition $(\dag')$ in
p.~\pageref{cdn:dagprime}.
Substituting (\ref{e1}) and (\ref{psie}) into (\ref{lax1})
gives a system of equations. The first three of them
are
\bea\label{eq1}
\a_i \dot x_i+2\b_i&=&0\,,\\
\label{eq2}
\dot\a_i+\a_i v_i+2\g_i&=&0\,,\\
\label{eq3}
\dot\b_i+v_i\b_i-\g_i\dot x_i+\a_i w_i&=&0\,.
\eea
Taking the $y$-derivative of the first equation and using the others, we get
(\ref{cm5}).

The equation (\ref{cm5}) is all what we are going to use below. Let us show
that it is valid for any meromorphic solution of the KP equation.
Namely: let $\tau(x,y,t)$ be a a holomorphic function of the variable $x$ in
some domain in $\bC$, depending smoothly on parameters $y$, $t$ and having
only simple zeros $x=x_i(y,t)$.
 
\begin{lem}
If the function
$u=-2\p_x^2\ln \tau(x,y,t)$ is a solution of the KP equation then
the equations (\ref{cm5}) hold.
\end{lem}
\emph{Proof.} For the proof of the theorem it is enough to substitute the Laurant
expansion of $u$ at $x(y,t)$ into the KP equation and consider
the coefficient in front of $(x-x_i(y,t))^{-3}$.

Next, let us show that equations (\ref{cm5}) are sufficient for the
existence of meromorphic wave solutions, i.e., solutions of the form
\beq\label{ps}
\psi(x,y,k)=
e^{kx+(k^2+b)y}\Biggl(1+\sum_{s=1}^{\infty}\xi_s(x,y)\,k^{-s}\Biggr)\,,
\eeq
where $b$ is a constant, $\xi_s$ are meromorphic functions, and the series
in parentheses is a formal power series in $k^{-1}$.
\begin{lem}\label{localsolv}
Suppose that equations (\ref{cm5}) for the zeros of $\tau(x,y)$ hold.
Then there exist meromorphic wave solutions of equation (\ref{lax1}) that
have simple poles at $x_i$ and are holomorphic everywhere else.
\end{lem}
\emph{Proof.}
Substituting (\ref{ps}) into (\ref{lax1}) gives a recurrent system of
equations
\beq\label{xis}
2\xi_{s+1}'=\dot\xi_s+(u+b)\xi_s-\xi_s''.
\eeq
Adding $b$ to $u$ does not change the coefficient $w_i$ in the expansion
(\ref{e1}), so the presense of $e^{by}$ in (\ref{ps}) has
no effect on the assertion of the lemma.
We are going to prove by induction that this system has meromorphic
solutions with simple poles at $x=x_i$\,.

Let us expand $\xi_s$ at $x=x_i$\,:
\beq\label{5}
\xi_s={r_s\over x-x_i}+r_{s0}+r_{s1}(x-x_i)+O((x-x_i)^2)\,,
\eeq
where we omit the index $i$ in the notation for the coefficients of this
expansion, since it suffices to look at a neighborhood of each $x_i$.
Suppose that $\xi_s$ are defined and equation (\ref{xis}) has a meromorphic solution.
Then the right-hand side of (\ref{xis}) has no residue at $x=x_i$, i.e.,
\beq\label{res}
\res_{x_i}\left(\p_y\xi_s+u\xi_s-\xi_s''\right)=\dot r_s+v_ir_s+2r_{s1}=0
\eeq
We need to show that the residue of the next equation vanishes also.
 From (\ref{xis}) it follows that the coefficients of the Laurent expansion for $\xi_{s+1}$
are equal to
\bea\label{6}
r_{s+1}&=&-\dot x_ir_s-2r_{s0}\,,
\\
\label{7}
2r_{s+1,1}&=&\dot r_{s0}-r_{s1}+w_ir_s+v_ir_{s0}\,.
\eea
These equations imply
$$
\dot r_{s+1}+v_ir_{s+1}+2r_{s+1,1}=-r_s(\ddot x_i-2w_i)-\dot x_i(\dot r_s-v_ir_ss+2r_{s1})=0\,,
$$
and the lemma is proved.

\paragraph{Step 2}
Let us now reintroduce $z$-dependence to $\tau$, so that it is
a function of $z+Ux\in\bC^n$ and $y$.
Our goal is to fix a \emph{translation-invariant} normalization of $\xi_s$
to define wave functions uniquely up to an $x$-independent factor.

We assume that $y$ runs over a small neighborhood $D$ of $0\in\bC$, and
let $\bC^{n*}:=\bC^n\times D$.
Identify $\bC^n$ with $\bC^n\times\{0\}\subset\bC^{n*}$, and
hence $U\in\bC^n$ with $(U,0)\in\bC^n\times\bC$ and $\Lambda$ with
$\Lambda\times\{0\}$.  A $\Lambda$-invariant subset of $\bC^{n*}$
will be regarded as a subset of $X^*:=X\times D=\bC^{n*}/\Lambda$.
Let $\Theta:=\{(z,y)\in X^*\mid \tau(z,y)=0\}$,
$\Theta_1:=\{(z,y)\in X^*\mid \tau(z,y)=\p_U\tau(z,y)=0\}$.
The singular locus
$\Sigma=\bigcap_{x\in\bC}(\Theta+Ux)=\bigcap_{x\in\bC}(\Theta_1+Ux)$
is a unique maximal $\bC U$-invariant subset of $\Theta_1$.
As we observed in Section~\ref{sec:zeroloc}, dividing
$\tau$ and $\tau_A$ by suitable $\p_U$-invariant functions,
we assume that $\Sigma$ and $\Sigma_A$ are of codimension${}\ge2$ in $X$.
Then $\Theta_1$ is also of codimension${}\ge2$ in $X$.
Let $Y_U=\langle\bC U\rangle$ be the Zariski closure of the group $\bC U$
in $X$.  Since it is a minimal $\bC U$-invariant closed subset
of $X^*$, $\Sigma$ and $\Sigma_A$ are $Y_U$-invariant, so that for any
$(z,y)\in X^*$ we have either
$Y_U\cap(\Sigma-(z,y))=\emptyset$ or $Y_U\subset\Sigma-(z,y)$.
The former is true outside a set of $(z,y)$ of codimension${}\ge2$ in $X$.

Let $\pi\colon\bC^{n*}\to X^*$ be the covering map, let
$\bC^d=\pi^{-1}(Y_U)^0$ be the connected component of $\pi^{-1}(Y_U)$
through the origin, and let $\Lambda_U:=\Lambda\cap\bC^d$.
Since $Y_U=\bC^d/\Lambda_U$ is compact,
$\Lambda_U$ is a lattice in $\bC^d$. Taking another vector subspace $H$
of $\bC^n$ such that $\bC^n=\bC^d\oplus H$,
we can write any $z\in\bC^n$ as $z=z'+z''$, where $z'\in\bC^d$ and
$z''\in H$.  Consider $\tau$ as a function of $z'\in\bC^d$ and $y\in D$
depending on parameters $z''\in H$.
The function $u(z,y)=-2\p_U^2\ln\tau$ is periodic with respect to
$\Lambda_U$ and, for each $(z'',y)$, has a double pole in $z'$ along
the divisor $\Theta^U(z'',y):=\left(\Theta-(z'',y)\right)\cap Y_U$.

\begin{lem}\label{qplem} Suppose the equation
\beq\label{rr}
\res_x\left(\p_y^2\ln \tau+2\left(\p_x^2\ln \tau\right)^2\right)=0
\eeq
for $\tau(Ux+z,y)$ holds, and let $\l_1$, \dots, $\l_d$ be
$\bC$-linearly independent vectors in $\Lambda_U$.  Then

(i) equation (\ref{lax1}) with the potential $u(Ux+z,y)$
has a wave solution of  the form $\psi=e^{kx+k^2y}\phi(Ux+z,y,k)$
such that the coefficients $\xi_s(z,y)$ of the formal series
\beq\label{psi2}
\phi(z,y,k)=e^{by}\Biggl(1+\sum_{s=1}^{\infty}\xi_s(z,y)\, k^{-s}\Biggr)
\eeq
are $(\l_1,\dots,\l_d)$-periodic meromorphic functions of $(z,y)\in\bC^{n*}$
with a simple pole along the divisor $\Theta^U$,
\beq\label{v1}
\xi_s(z+\l_i,y)=\xi_s(z,y)={\tau_s(z,y)\over \tau(z,y)}\,,\quad i=1,\dots,d\,;
\eeq

(ii) $\phi(z,y,k)$ is unique up to a factor which is $\p_U$-invariant
and holomorphic in $z$, i.e., if $\phi$ and $\phi_1$ are two solutions,
then we have
\beq\label{v2}
\phi_1(z,y,k)=\phi(z,y,k)\rho(z'',k)\,.
\eeq
\end{lem}
\emph{Proof.}
Let us temporarily modify formula (\ref{psi2}) for $\phi$\,:
$$
\phi(z,y,k)=e^{by+\sum_{s=1}^\infty b_s(y)k^{-s}}
\Biggl(1+\sum_{s=1}^{\infty}\xi_s(z,y)\, k^{-s}\Biggr)\,,
\eqno(\ref{psi2}')
$$
where $b_s(y)=b_s(z'',y)$ are functions of $y$ and $z''$, i.e.,
they are independent of $z'$, such that $b_s(0)=b_s(z'',0)=0$.
The factor $e^{\sum_{s=1}^\infty b_s(y)k^{-s}}$ can later be absorbed
into $1+\sum\xi_sk^{-s}$, so it is redundant, and harmless.

Substituting $\psi=e^{kx+k^2y}\phi$ and $(\ref{psi2}')$ into
equation (\ref{lax1}), we find the recursion formulas
\beq\label{xis1}
2\p_U\xi_{s+1}=(\p_y-\p_U^2+(u+b))\xi_s+\sum_{i=1}^s b_i'\xi_{s-i}\,,
\quad s=0,\ 1,\dots,
\eeq
where we set $\xi_0=1$.
The first equation $2\p_U\xi_1=u+b$ can be solved explicitly:
\beq\label{v5}
\xi_1=-\p_U\ln \tau +(l_1,z)\,,
\eeq
for a linear form $(l_1,\cdot)$ on $\bC^d$, and $b=2(l_1,U)$.
The periodicity condition (\ref{v1}) for $s=1$ is satisfied if and only if
\beq\label{v6}
(l_1,\l_i)=\p_U\ln \tau(z+\l_i,y)-\p_U\ln \tau(z,y)=
a_{\l_i}\cdot U\,,\quad i=1,\dots,d,
\eeq
where the last equality follows from (\ref{monod}).
Since $\l_1$, \dots, $\l_d$ are linearly independent vectors in $\bC^d$,
this determines the linear form $(l_1,\cdot)$ uniquely.
The form $(l_1,\cdot)$ is independent of $z$ and $y$ since
the right-hand side of (\ref{v6}) is, so $b=(l_1,U)$ is a constant.

Suppose we have $(\l_1,\dots,\l_d)$-periodic functions
$\xi_1$, \dots, $\xi_r$, constant $b$, and functions
$b_1(y)$, \dots, $b_{r-1}(y)$ of $y$ solving equations (\ref{xis1})
for $s<r$, and consider the equation for $s=r$, with tentative choice of
$b_r\equiv0$.

Equation (\ref{rr}) implies equations (\ref{cm5}), which, as seen in
Lemma~\ref{localsolv}, are sufficient for the local solvability of
(\ref{xis}), and hence (\ref{xis1}), away from the locus $\Theta_1$ of
multiple zeros of $\tau$ as a function of $x$.
This set does not contain a $\p_U$-invariant line away from $\Sigma$
which is of codimension${}\ge2$. 
Therefore, the sheaf $\V_0$ of $\p_U$-invariant
meromorphic functions on $\bC^{n*}\setminus \Theta^{\,U}_1$ with poles
along the divisor $\Theta^{\,U}$
coincides with the sheaf of $\p_U$-invariant holomorphic functions.
This implies the vanishing of $H^1(\bC^{n*}\setminus\Theta^U_1,\V_0)$
and the existence of global meromorphic solutions $\xi_s$ of (\ref{xis1}),
with a simple pole along the divisor $\Theta^{\,U}$
(see details in \cite{shiota, ac}).

Let $\xi_{r+1}^0$ be a solution, not necessarily periodic, of (\ref{xis1})
for $s=r$ with $b_r\equiv0$.  Then (\ref{xis1})
implies that $\xi_{r+1}^0(z+\l_i,y)-\xi_{r+1}^0(z,y)$, $i=1$,~\dots, $d$,
are constant in the $U$-direction, hence they are constant on each
translate of $Y_U$.  Let $(l_{r+1},\cdot)$ be a linear form on $\bC^d$,
depending on $y$ and $z''$, such that
$$
(l_{r+1},\l_i)=\xi_{r+1}^0(z+\l_i,y)-\xi_{r+1}^0(z,y)\,,\quad i=1,\dots, d\,,
$$
hold. Then
\beq\label{v7}
\xi_{r+1}(z,y)=\xi_{r+1}^0(z,y)+(l_{r+1},z)\quad\hbox{and}\quad
b_r(y)=2\int_0^y(l_{r+1}(y'),U)dy',
\eeq
together with the previously chosen $\xi_s$ ($s\le r$) and $b_s$ ($s<r$),
give a $(\l_1,\dots,\l_d)$-periodic solution of (\ref{xis1}) for $s=r$.
This completes the induction step, proving (i) except for the latter
half of (\ref{v1}), which is obvious if we compare the poles of $\xi_s$
and the zeros of $\tau$.

In each step of this construction, equation (\ref{xis1}) determines
$\xi_{s+1}$ uniquely up to an additive constant $c_{s+1}(z'')$
depending on $z''$.
Indeed, the constant may depend on $y$, but the effect of this
$y$-dependence will be cancelled by $b_{s+1}$ to be chosen in the
next step, so we assume the $c_{s+1}(z'')$ is independent of $y$.
Adding $c_{s+1}(z'')$ to $\xi_{s+1}(z,y)$ will affect the later steps
of construction, but in terms of $\phi$ all the necessary changes
can be done just by multiplying it by $1+c_{s+1}(z'')k^{-s-1}$.
This proves (ii), with
$\rho(z'',k)=\prod_s(1+c_{s+1}(z'')k^{-s-1})$\,.
The lemma is thus proven.


\section{Commuting differential operators}\label{sec:commDO}

In this section, using the wave function $\psi$ we show the existence
of sufficiently many commuting differential operators, to obtain the
curve $\G$. From the point of view of the KP hierarchy, this amounts to
showing the finite dimensionality of the orbit.  There can be several
approaches. For instance, given our specific form of quasiperiodicity
condition, i.e.,
that $\tau$ is of the form $\tau(Ux+z,y)$ and is periodic in $z$
with periods $\l_1$, \dots, $\l_d$,
knowing its zero locus is enough to recover $\tau$, and hence
$u$. Denote by $\Theta_0$ the space of ample divisors on $X$ which
belong to the given polarization. Then equations (\ref{cm5}) may be
seen as a dynamical system on a subset of the tangent bundle $T(\Theta_0)$
of $\Theta_0$.
This set is finite dimensional, and we can realize the
whole KP flows as commuting flows on this space, so the whole KP orbit
must be finite dimensional.  Rather than following this argument,
here we give a proof by showing the finite dimensionality of
certain space, to which we map an infinite sequence of differential
operators, thus showing that sufficiently many linear combinations of
the operators belong to the kernel of the map. We then identify this
kernel with a space of commuting differential operators.

First define a pseudo-differential operator
\beq\label{LL}
\L=\p_x+\sum_{s=1}^{\infty} w_s(z,y)\p_x^{-s}
\eeq
by
\beq\label{kk}
\L(Ux+z,\p_x)\,\psi=k\,\psi\,,
\eeq
or equivalently
\beq\label{sandwich}
\L(Ux+z,\p_x)=\Phi\p_x\Phi^{-1}\,,
\eeq
where
\beq\label{w-op}
\Phi=1+\sum_{s=1}^{\infty}\xi_s(Ux+z,y)\,\p_x^{-s}
\eeq
if $\phi=e^{-(kx+k^2y)}\psi$ is given by (\ref{psi2}).
So $\L$ is determined uniquely by $\psi$, and the ambiguity (\ref{v2})
in defining $\psi$ does not affect $\L$, so it is determined
by $u=-2\p_x\ln\tau$ and the choice of vectors $\l_1$, \dots, $\l_d$.
Since $u$ is $\Lambda$-periodic, so is $\L$\,; the coefficients
$w_s(Z,y)$ of $\L$ are meromorphic functions on $X^*$ with poles along
the divisor $\Theta$.

Consider now the differential parts of the pseudo-differential operators
$\L^m$, namely, let $\L^m_+$ be the differential operator such that
$\L^m_-:=\L^m-\L^m_+=F_m\p^{-1}+F_m^1\p^{-2}+F_m^2\p^{-3}+O(\p^{-4})$.
Here we denote $\p_x$ by $\p$ for simplicity.
The leading coefficient $F_m$ of $\L^m_-$ is the residue of $\L^m$\,,
\beq\label{res1}
F_m=\res_\p\L^m\,,
\eeq
and
\beq\label{res12}
F_m^i=\res_\p(\L^m\p^i)\,.
\eeq
 From the construction of $\L$ it follows that $[\p_y-\p^2_x+u, \L^n]=0$. Hence
\beq\label{lax}
[\p_y-\p_x^2+u,\L^m_+]=-[\p_y-\p_x^2+u, \L^m_-]=2\p_x F_m\,.
\eeq
The vanishing of the coefficients of $\p^{-1}$ and $\p^{-2}$ in the
middle member of this equality
implies the equations
\bea\label{van}
2\p_x F_m^1&=&-\p_x^2 F_m+\p_yF_m.
\\
\label{van2}
2\p_x F_m^2&=&F_m u_x-\p_x^2 F_m^1+\p_yF_m^1.
\eea
The functions $F_m, F_m^i$ are differential polynomials in the coefficients $w_s$ of $\L$.
Hence, they  are meromorphic functions on $X$.
\begin{lem} The abelian functions $F_m$ have at most the second order pole
on the divisor $\Theta$.
\end{lem}
\emph{Proof.}
The $m$th KP flow on $\Phi$ is defined by
\beq\label{kpwave}
\p_{t_m}(\Phi)=\L^m_+\Phi-\Phi\p^m=-\L^m_-\Phi\,.
\eeq
Comparing the coefficients of $\p^{-1}$ on both sides of (\ref{kpwave})
and using (\ref{v5}), we obtain
\beq\label{res-kpwave}
F_m=\p_x\p_{t_m}\ln\tau\,.
\eeq
So if we admit that the higher KP flows preserves the regularity of
$\tau$, the assertion follows immediately from (\ref{res-kpwave}).

Alternatively, by constructing the adjoint wave function, we can see
that the 0th order pseudodifferential operator $\Phi^{-1}$ can be
written in the form
\beq\label{w-op-1}
\Phi^{-1}=1+\sum_{s=1}^\infty\p^{-s}\circ\xi^*_s(Ux+z,y)
\eeq
for some meromorphic functions $\xi^*_s$ having a simple pole along
$\Theta$.  Using (\ref{sandwich}), (\ref{w-op}) and (\ref{w-op-1}) we have
$$
\L^m=\Phi\p^m\Phi^{-1}=\sum_{r,s=0}^\infty\xi_r\p^{m-r-s}\circ\xi^*_s\,,
$$
where we set $\xi_0=\xi^*_0=1$. Since $\xi_r\p^{m-r-s}\circ\xi^*_s$
does not yield negative order terms in $\p$ if $m-r-s\ge0$,
this implies
\beq\label{minuspart}
\L^m_-=\sum_{r+s>m}\xi_r\p^{m-r-s}\circ\xi^*_s=
\sum_{r+s=m+1}\xi_r\xi^*_s\p^{-1}+O(\p^{-2})\,,
\eeq
hence $F_m=\sum_{r+s=m+1}\xi_r\xi^*_s$\,, proving the lemma.

\paragraph{Important remark}
In \cite{kr-schot} this statement was crucial for
the proof of the existence of commuting differential operators
associated with $u$. Namely, it implies that for all but a finite number
of positive integers $n$ there exist constants $c_{n,i}$ such that
\beq\label{f1}
F_n(z,y)+\sum_{i=0}^{n-1} c_{n,i}F_i(z,y)=0\,,
\eeq
hence (\ref{lax}) would imply that the corresponding linear combinations
$L_n:=\L_+^n+\sum c_{n,i}\L_+^i$ commutes with $P:=\p_y-\p_x^2+u$.
Not so: since these constants $c_{n,i}$ might depend on $y$,
we might not have $[P,L_n]=0$, and we cannot immediately make the
next step and claim the existence of commuting operators (!).

So our next goal is to show that these constants in fact are
$y$-independent.
For that let us consider the functions $F_m^1$.
Equation (\ref{van}) (or (\ref{minuspart})) implies that they have at most
the third order pole on the divisor $\Theta$\,.
Moreover, if we expand $F_m^1$ near $\Theta$\,,
\beq\label{F1expan}
F_n^1={f_n^3\over \tau^3}+{f_n^2\over \tau^2}+{f_n^1\over \tau}+O(1)\,,
\eeq
and use (\ref{res-kpwave}) so that $F_n$ is of the form
\beq\label{sh1}
F_n=\p_U\left({q_n\over \tau}+O(1)\right)=-{q_n\p_U\tau\over\tau^2}+
{\p_Uq_n\over\tau}+O(1)\,,
\eeq
then (\ref{van}) implies
\beq\label{sh2}
f_n^3=-q_n(\p_U\tau)^2,\quad f_n^2=0\,.
\eeq
Let $\{F_\a^1\mid\a\in A\}$, for finite set $A$, be a basis of the
space $\F(y)$ spanned by $\{F_m^1\}$. Then for
all $n\notin A$ there exist constants $c_{n,\a}(y)$ such that
\beq\label{sh3}
F_n^1(z,y)=\sum_{\a\in A} c_{n,\a}(y)F_\a^1(z,y)\,.
\eeq
Due to (\ref{sh2}) it is equivalent to the equations
\bea\label{sh4}
q_n(z,y)&=&\sum_{\a} c_{n,\a}(y)q_\a(z,y)\,,
\\
\label{sh5}
f_n^1(z,y)&=&\sum_{\a} c_{n,\a}(y)f_\a^1(z,y)\,.
\eea
 From equation (\ref{van})
we get
\beq\label{sh6}
\sum_{\a} (\p_yc_{n,\a})q_\a(z,y)=0\,.
\eeq
Taking a linear combination of (\ref{van2}) we
get
\beq
\label{sh7}
2\p_x\Biggl(F_n^2-\sum_{\a}c_{n,\a}F_\a^2\Biggr)=\sum_{\a} (\p_yc_{n,\a}){f_\a^1\over \tau}+
O(1)\,.
\eeq
The left-hand side has no ``residue" on $\Theta$, and that implies
the equation
\beq\label{sh8}
\sum_{\a} (\p_yc_{n,\a})
f_\a^1=0\,.
\eeq
Equations (\ref{sh4}) and (\ref{sh8}) are equivalent to
\beq\label{sh9}
\sum_{\a} (\p_yc_{n,\a})F_\a^1=0\,.
\eeq
By definition the functions $F_\a^1$ are linearly independent.
Therefore $c_{n,\a}$ are $y$-independent and we can proceed
as in \cite{kr-schot}.  Let us sketch the rest of the proof.

Now we have sufficiently many ordinary differential operators $L_n$,
one for each $n\gg0$, satisfying
\beq\label{lax2}
[P,L_n]=0\,.
\eeq
Although $P=\p_y-\p_x^2+u$ is a partial differential operator,
this suffices to conclude that $L_n$'s commute with each other.
Indeed, (\ref{lax2}) implies that $\tilde\psi:=L_n\psi$ satisfies
$P\tilde\psi=0$. Since $L_n$ is a linear combination of $\L_+^k$'s,
one observes that $\tilde\psi$ also satisfies the same periodicity
conditions as $\psi$, so by part (ii) of Lemma~\ref{qplem} it is
equal to $\psi$ up to a $\p_U$-independent factor. This implies
that the $L_n$'s commute with $\L$ and with each other.

The coefficients $c_{n,\a}$ of linear combinations give the
directions of trivial KP time evolutions, and the
Laurent coefficients of the polar part at $P$ of the
corresponding functions $f_n(\z)$ on the curve $\G$. Now let us find
\emph{all} the Laurent coefficients, not just the polar part, of
$f_n(\z)$.
Since $L_n$ commutes with
$\L$ and the latter is a first order operator, we can write
$L_n$ as a constant coefficient Laurent series in $\L^{-1}$:
$$
L_n=\sum_{j=-\infty}^n c_{n,j}\L^j,
$$
then $f_n(\z)=\sum_{j=-n}^\infty c_{n,-j}\z^j\in\bC((\z))$ is
the desired Laurent series for $f_n(\z)$.  The coefficients $c_{n,j}$
are constant in $z$ since they are periodic holomorphic functions.
Hence the curve $\G$ is constant in $z$.  They are also constant in $y$
since KP flows do not deform a spectral curve.


\end{document}